# Reconstruction of tokamak plasma safety factor profile using deep learning

Xishuo Wei[1]*, Ge Dong [2,3]*, Shuying Sun[4]*, William Tang[2], Zhihong Lin[1], Hongfei Du[3]

1. University of California, Irvine, CA
2. Princeton Plasma Physics Laboratory, NJ
3. Energy Singularity, Shanghai, China
4. Fusion Simulation Center, Peking University, Beijing, China

dongge@energysingularity.cn
*These authors contributed equally to this work

Abstract
In tokamak operations, accurate equilibrium reconstruction is essential for reliable real-time control and realistic post-shot instability analysis. The safety factor (q) profile defines the magnetic field line pitch angle, which is the central element in equilibrium reconstruction. The motional Stark effect (MSE) diagnostic has been a standard measurement for the magnetic field line pitch angle in tokamaks that are equipped with neutral beams. However, the MSE data are not always available due to experimental constraints, especially in future devices without neutral beams. Here we develop a deep learning-based surrogate model of the gyrokinetic toroidal code for q profile reconstruction (SGTC-QR) that can reconstruct the q profile with the measurements without MSE to mimic the traditional equilibrium reconstruction with the MSE constraint. The model demonstrates promising performance, and the sub-millisecond inference time is compatible with the real-time plasma control system.

## I. Introduction

Accurate equilibrium reconstructions provide crucial support for successful plasma operation and control for fusion devices such as DIII-D or ITER. The commonly used reconstruction method, like EFIT[L. L. Lao, NF1985; L. L. Lao, NF1990], uses the iterative algorithm to minimize the total error between computed values and measured values of some key parameters. The final reconstruction and the subsequent physical analysis can depend sensitively on the capability of measurements. For example, there are several versions of EFIT, among which EFIT01 and EFIT02 are widely used. EFIT02 takes advantage of the Motional Stark effect (MSE) diagnostic [Wróblewski 1992], and by combining the external magnetic measurements, the MSE measurement enables the accurate reconstruction of the q profile. Figure 1 shows a comparison of the simulation results of internal kink mode using the reconstructed equilibrium by the EFIT01 and EFIT02 equilibria for the same shot (#168973) at an identical time (4676 ms). At the identical time in the same shot, using the same set of simulation control parameters (such as the time step size and grid number), only equilibrium from the reconstruction with the MSE constraint, i.e. EFIT02 leads to unstable kink modes. The direct reason for this distinct difference in the current driven kink instability is that the q profile from EFIT02 is less than 1 near the axis, while the minimum q value from EFIT01 is larger than 1. Therefore, the reconstruction from EFIT02 (or other EFIT versions using more constraints) is preferable to EFIT01 to obtain an accurate physics analysis. A potential limitation of EFIT02 is that the reconstruction relies heavily on the measurement of the magnetic pitch angle and radial electric field using the MSE diagnostic which in turn is strongly dependent on the neutral beam settings in each individual experiment. [C. T. Holcomb 2006; C. T. Holcomb 2008]. However, there is a significant portion of plasma operations/shots where the MSE diagnostic is unavailable, or only available for limited periods of time during operation for the real-time EFIT ("rtEFIT") equilibrium reconstruction. To address this long-standing problem, it is desirable to build a statistical model based on deep learning techniques which use available diagnostic signals to reconstruct the q profile as accurately as EFIT02.

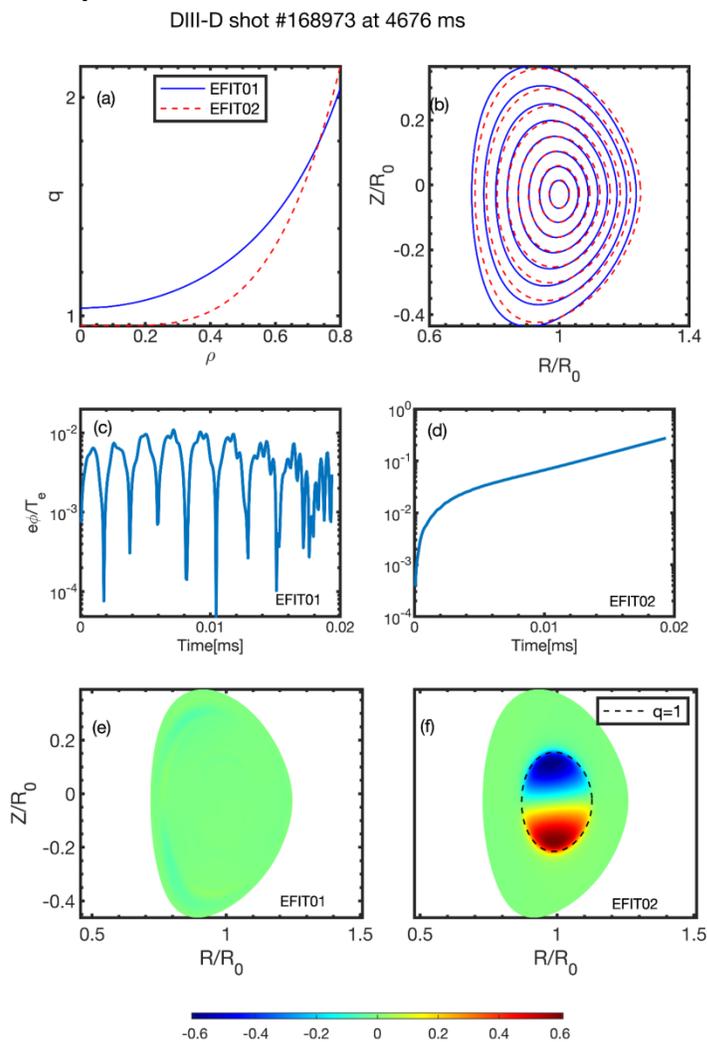

Figure 1. Comparison of GTC simulation result using EFIT01 equilibrium reconstruction in the left panels and EFIT02 equilibrium reconstruction in the right panels for the same shot (#168973) at the identical time (4676 ms). The q profile (panel a) and the shape of flux

surfaces(panel b) are compared. (c) and (d) are the time history of the m=1, n=1 mode. (e) and (f) are the poloidal mode structure of the perturbed electrostatic potential. The same color scale is used in (e) and (f), and (e) only exhibits the noise level fluctuations.

The information from instability measurements can significantly enhance equilibrium reconstructions. For example, the plasma rotation profile combined with distinguishing features of current driven instabilities and Alfven Avalanches from the spectrogram of magnetic signals can be correlated with constraints in the q profile, such as the minimum q and radial location of the q=1 surface. Traditional methods like the multi-step EFIT reconstruction based on measured plasma instabilities such as that described in [D. Spong 2012] require time-costly manual investigations for post-shot analysis. The deep learning based model trained with inputs including the measured plasma perturbation signals is capable of automatically combining and extracting useful information from the equilibrium and instability measurements and predicting the q profile in the order of or shorter than 1 millisecond.

In this work, we present the results of building and testing a statistical model based on a deep neural network, which takes as inputs the measured plasma perturbation signals, combined with the real-time EFIT results without MSE constraints, and outputs the q profile. The model is trained on about 12000 DIII-D EFIT02 reconstruction data, The output q profile is close to EFIT02 results, which shows the MSE constraints are implicitly included in the trained model. The sensitivity test on input parameters is carried out, showing the relative importance of each parameter to the accurate q reconstruction. This model can act as the SGTC q-Reconstruction module (SGTC-QR). The outputs of SGTC-QR provide more accurate plasma equilibrium information to enhance instability analysis and predictive models in the plasma control system (PCS), such as the SGTC MHD instability simulator[Dong 2021] and FRNN disruption predictors[J. Kates-Harbeck 2019]. The inference time of the pre-trained surrogate models from SGTC-QR will be compatible with the requirements of the real-time plasma control system. Previously, the machine learning and deep learning methods has been introduced to solve the equilibrium reconstruction, such as the works to solve Grad-Shafranov equation[van Milligen 1995; S. Joung 2019], and the recent machine learning implementation in the EFIT-AI framework[L. L. Lao 2022]. The previous works are focused on finding the surrogate model of a known physical equation, i.e., the Grad-Shafranov equation, with given constraints. While in this work, we investigate whether it is possible to obtain a high-quality reconstruction without the important constraint, the MSE signal. In other words, the machine learning algorithm tries to recover the information contained in the MSE measurements from other measurements, and there is no physical model for this purpose. In the future, the recovery of other measurements can also be studied. A complete equilibrium reconstruction, including the flux surface shape, will also be investigated.

In Section 2, we introduce the building and training details of SGTC-QR. The testing results of SGTC-QR and an example of physical analysis based on the predicted q profile are presented in Section 3. The conclusion is drawn in Section 4.

**II. Workflow and data for q profile reconstruction**

The workflow of SGTC-QR is shown in Figure 2. The real-time measurements will provide some global equilibrium parameters such as the total current, the stored energy, and MSE (if present) and the perturbation signal such as the ECE and magnetic perturbations. Through the offline or real-time EFIT, the 2D equilibrium is reconstructed, including the flux surface shape, q-profile, current profile, pressure profile, etc. When the MSE measurement is not present, we can incorporate all available equilibrium and perturbation signals in the SGTC-QR module to generate the accurate q profile and the 2D flux surface shape can also be regenerated in principle. Then these equilibrium profiles are taken as input by the stability models or predictors like SGTC or FRNN. The stability estimations act as feedback to the control systems to modulate the instabilities.

The architecture of SGTC-QR is shown in the lower panel of Figure 2. We select 16 real-time signals (listed in Table 1) as input of SGTC-QR. For each time step, the electron density profile $n_e$ and electron temperature profile $T_e$ are 1-D functions on the Nr grid points of the normalized square root of toroidal flux $\rho$, the magnetic perturbations are a function of frequency $\omega$, and the other signals are scalar variables. The $n_e$ and $T_e$ are combined to an Nr * 2 2d feature and fed into a set of Nc1 convolutional layers. $\delta B$ act as a 1d feature of size $N\omega$, and feed into another set of Nc2 convolutional layers. The scalar variables are fed in a set of Nf1 fully connected layers. The output of the convolutional layers and the fully connected layers are stacked together to go through Nf2 fully connected layers. The final output layer has the size of Nr, which corresponds to the q profile on grid points of $\rho$.

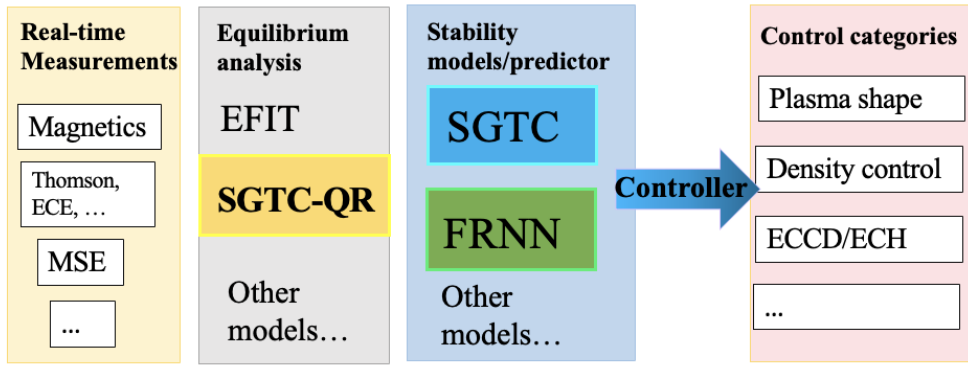

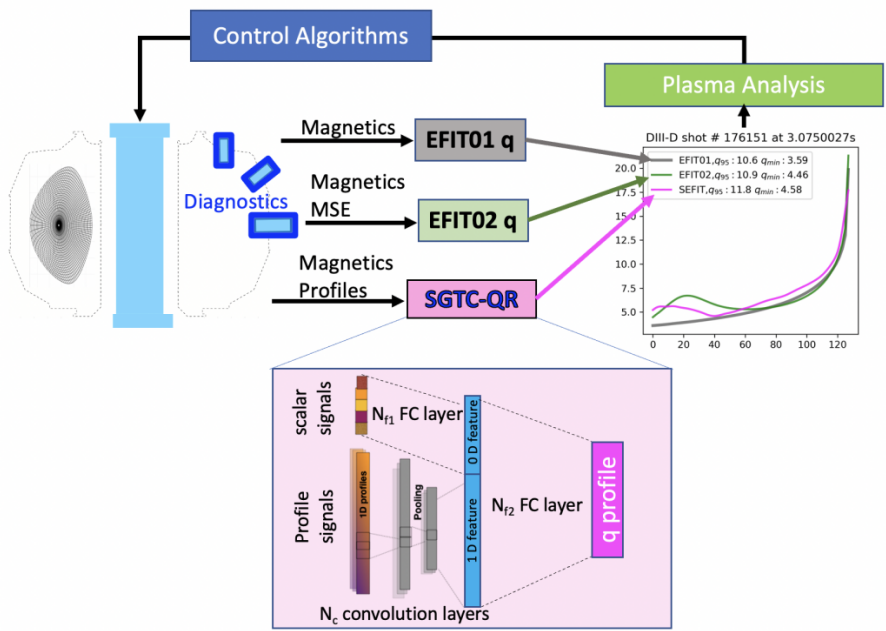

Figure 2. Workflow of SGTC-QR. Upper panel shows the Workflow of SGTC. Lower panel shows the workflow of SGTC-QR in detail.

The data for training and testing SGTC-QR are randomly selected from the DIII-D shots #125000-#180844. In total, 13491 shots are picked, and 4521 of them are disruptive. 6780 shots are chosen to compose the training data set, with 2267 of them being disruptive; 3339 for the validation data set with 1122 of them being disruptive and another 3372 for the testing data set with 1132 of them being disruptive. The equilibrium slices are evenly spaced and selected every 25ms from these shots, as a result, each of them typically contains 110 time-sliced equilibriums on average. The models are trained and tested in the PyTorch framework. The ensemble method is used to reduce stochastic errors. We build 40 different networks, with different learning rate schedulers, and different hidden layer depths and widths. The mean-square-error between the output q profile and the EFIT02 q profile is taken as the loss function. Note that the training does not involve any output from EFIT01. The training process uses the stochastic gradient descent (SGD) optimizer, with a minibatch size of 4. The 40 networks are trained parallelly on the same training data set. The whole training takes less than 1 hour for one epoch when training using 1 node with 128 processors and 4 GPUs on NERSC's Perlmutter cluster. The training process would be terminated when the validation loss no longer drops in 1000 epochs to avoid overfitting.

| Signal descriptions | Symbols | Usage in SGTC-QR | Shot number |
|---|---|---|---|
| Safety factor profile from EFIT02 | $q(\rho)$ | Output | *12000 equilibriums from shot # 125000–180844* |
| Safety factor profile from EFIT01 | | Not used | |
| Electron temperature profile | $T_e(\rho)$ | Input | |
| Electron density profile | $n_e(\rho)$ | | |
| Mirnov spectrogram | $\delta B(\omega)$ | | |
| Internal Inductance | $L_i$ | | |
| Plasma density | $n_{e0}$ | | |
| Safety factor at 95% radial domain from EFIT01 | $q_{95}$ | | |
| Plasma current | $I_p$ | | |
| Ratio of thermal to magnetic pressure | $\beta$ | | |
| Plasma stored energy | $W_{MHD}$ | | |
| Input beam power | $P_{in}$ | | |
| Input beam torque | $T_{in}$ | | |
| Pressure profile pedestal height | $P_{ped}$ | | |
| Temperature profile pedestal height | $T_{ped}$ | | |
| Density profile pedestal height | $n_{ped}$ | | |
| Temperature profile pedestal width | $\Delta_T$ | | |
| Density profile pedestal width | $\Delta_n$ | | |

Table 1. Data used in training, validation, and testing of SGTC-QR. $\rho$ is the normalized toroidal flux that defines the plasma radial domain, and $\omega$ is the frequency in the spectrogram.

## III. Results

After the training is complete, 5 networks with the best validation accuracy are chosen, and the average value of their outputs is taken as the prediction for the q profile. The average prediction time is less than 1 millisecond, meaning this model is highly compatible with the real time plasma control systems. Figure 3 shows the comparison of the reconstructed q profile of the SGTC-QR model and that from EFIT01 and EFIT02. The average mean-square-error of SGTC-QR is 0.165, lower than the average mean-square-error of EFIT01, 0.232. The median mean-square-error of SGTC-QR (0.064) is also lower than that of EFIT01(0.080). The distribution of the mean-square-error of the q profile is shown in Figure 3, from which we see SGTC-QR tends to show smaller errors. This means in general the SGTC-QR reconstructed q profile is more accurate than EFIT01. In Figure 4 we compare the two important features, the q95, and qmin, which affect the physics near the edge and core in tokamaks, between reconstructed q profiles. Here we take the EFIT02 results as references, the deviation of the points from the y=x line shows the difference of reconstructed q95 or qmin with EFIT02 results. From the upper panel of Figure 4, we see the q95 difference of SGTC-QR with EFIT02 is comparable to that of EFIT01. In the lower panel of Figure 4, the minimum values of SGTC reconstructed q profiles are shown close to EFIT02 than EFIT01, which potentially means the reconstruction in the core region is more accurate. This is also demonstrated by the radial error distribution in Figure 5. The errors in each of the 128 radial points are calculated for all shots in the test dataset. It clearly shows that both the average and medial of the errors from SGTC-QR reconstruction is lower than EFIT01 when $\rho < 0.8$. The reason for the large error near the edge lies in two aspects: q95 is taken as a constraint in EFIT01 and EFIT02 reconstruction, which enforces the q95 value to be close to the measured one during the iteration of EFIT calculation. While SGTC-QR only takes q95 as an input parameter, without regarding it as a constraint. The second part of the reason is that the q value at the separatrix is infinite, and the reconstruction in EFIT02 (and of course also in EFIT01) is inaccurate near the separatrix. So it is very difficult for the deep learning method to find the pattern of the q profile from EFIT02 near the separatrix. In fact, we can see the increase of errors when $\rho$ approaches 1, especially at the last 3 radial points. That is why we did not plot the last 3 radial points in Figure 5. If we only care about the physics except the region for the last 3 points, the effective error of the q profile from SGTC-QR would be even smaller than that shown in Figure 3.

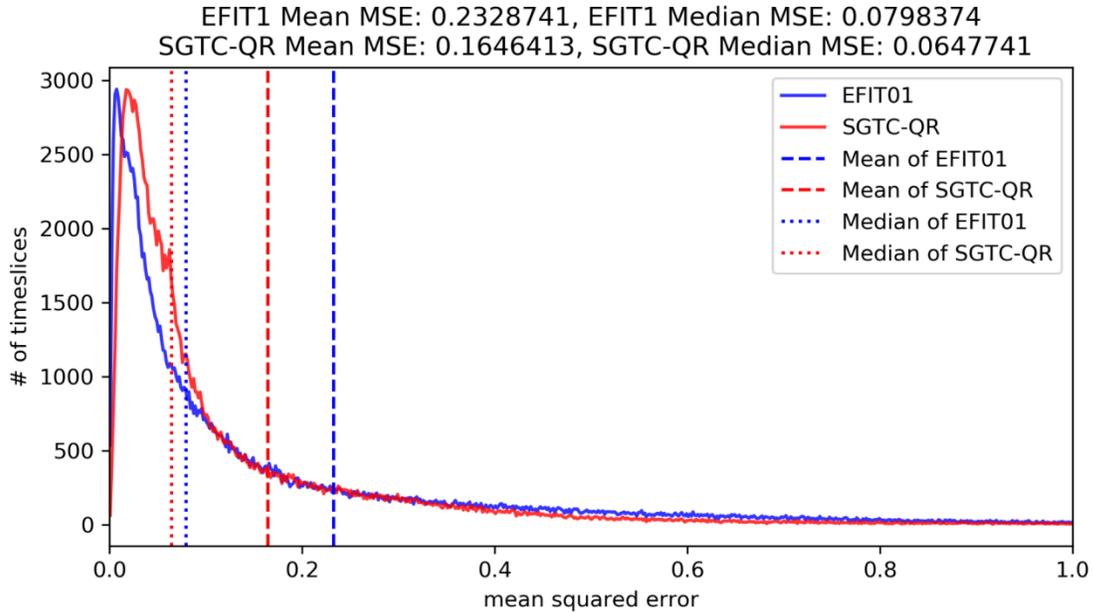

Figure 3. Comparison of the mean squared error of EFIT01 q profile and SGTC-QR q profile, using EFIT02 q profile as the true value. Data are shown for the test dataset with 164782 time slices from 3372 shots.

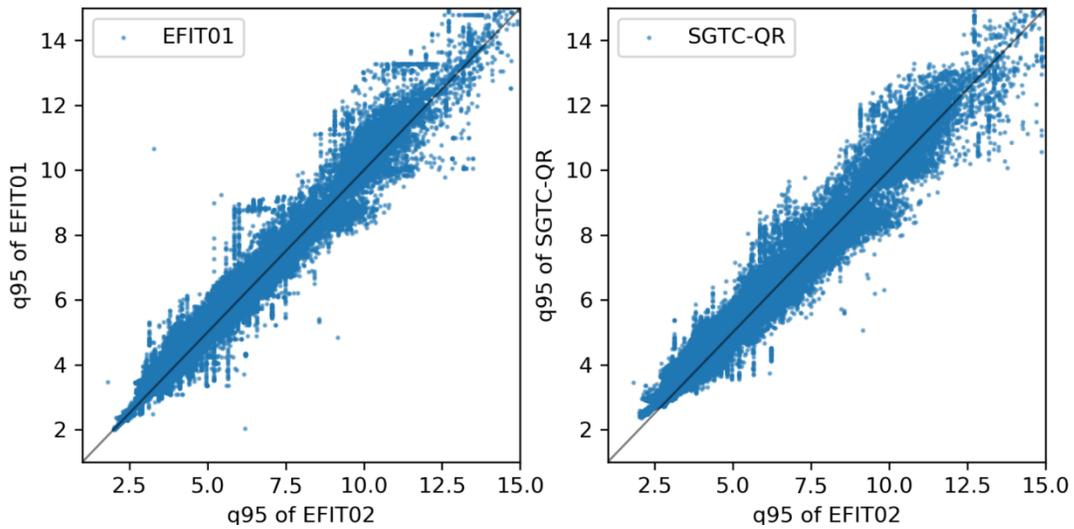

Figure 4-1. Comparison of the q95 given by EFIT01 and SGTC-QR, using q95 from EFIT02 as the truth value. Data are shown for the test dataset with 164782 time slices from 3372 shots.

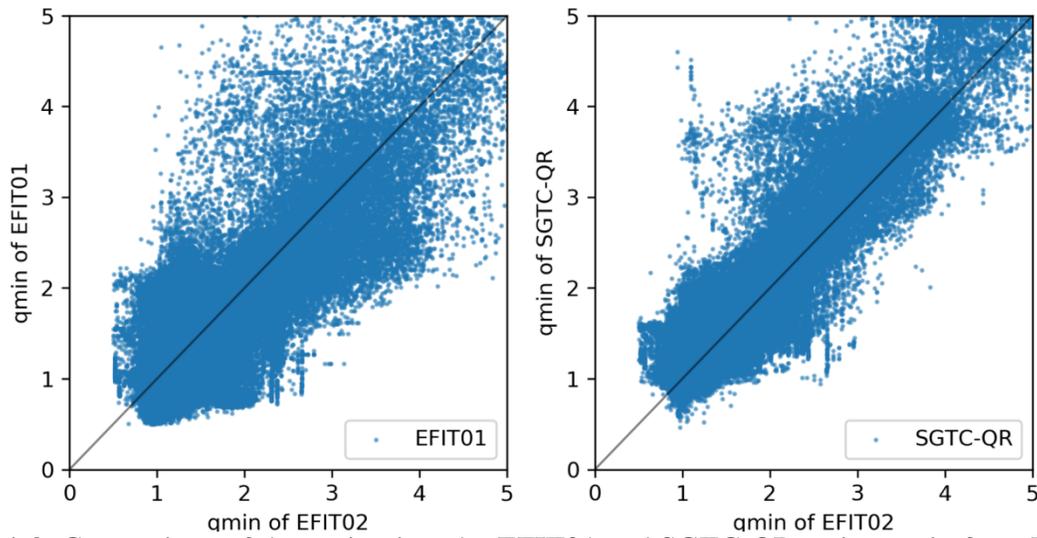

Figure 4-2. Comparison of the qmin given by EFIT01 and SGTC-QR, using qmin from EFIT02 as the truth value. Data are shown for the test dataset with 164782 time slices from 3372 shots.

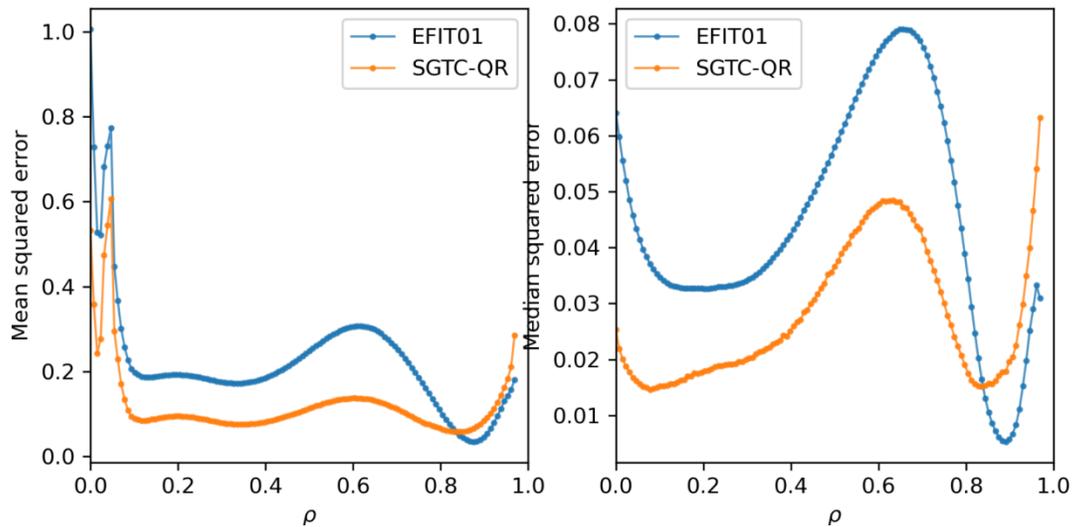

Figure 5. Comparison of mean-square-error of EFIT01 and SGTC-QR at each radial point. (Last 3 radial grid points are not shown)

From the above results, we conclude that the reconstructed q profile outperforms the EFIT01 and successfully incorporates the MSE constraint implicitly. The underlying reason for the successful reconstruction is revealed through a sensitivity study. In this study, we contaminate one of the inputs by directly setting it to 0 each time, and measure the prediction accuracy on the testing data set. The increase of the median and average mean-square-error value from the original one reflects the relative importance of each input parameter. The result is shown in

Figure 6, and the title of each line stands for the input contaminated. The lines are arranged in the order of decreasing average mean-square-error, and all the errors are normalized so that the median and average mean-square-error of the original SGTC-QR is 1. We also show the error from the EFIT01 q profile reconstruction in the figure. Figure 6 shows the most important input for successful q reconstruction is the plasma current, the internal inductance, and the normalized beta. The importance of these factors to the q profile makes physical sense because they directly affect the current and pressure profile, and directly determine the solution of the Grad-Shafranov equation, and subsequently the q profile. Notably, the q95 value and the magnetic fluctuation signal do not make much difference in the improvement of the reconstruction. And this also indicates that the SGTC-QR does not take the information from the q95 factor efficiently, and it potentially causes an increase in the error in the edge region.

In Figure 7, we show the q profile reconstruction for two typical cases. The line for SGTC-QR is from the one model with the best validation score among all 40 models. The left panel is from the benchmark case for internal kink mode [Brochard 2022], in which The EFIT01 reconstruction is invalid. We see both EFIT02 and SGTC-QR show normal shear, and the q profile near the axis is very similar, which is important for correct kink drive. The right panel is from the benchmark case for RSAE [W. Deng 2012]. The EFIT02 q profile shows normal shear. But the SGTC-QR shows reversed shear, with $q_{min} \approx 3.1$. The inconsistency between the two q profiles does not mean the failure of the machine learning model. More accurate analysis with human calibration indicates the q-profile should be reversed sheared with $q_{min}=3.18$. In this case, it is the EFIT02 q profile that deviates from the correct data distribution. And by learning on the whole training dataset, the robust machine learning model successfully predicts the correct q profile according to the learned data distribution. And in this sense, the SGTC-QR has the potential to warn and correct the q profile if the q profile reconstruction in the test dataset is inaccurate and deviates from the typical data distribution.

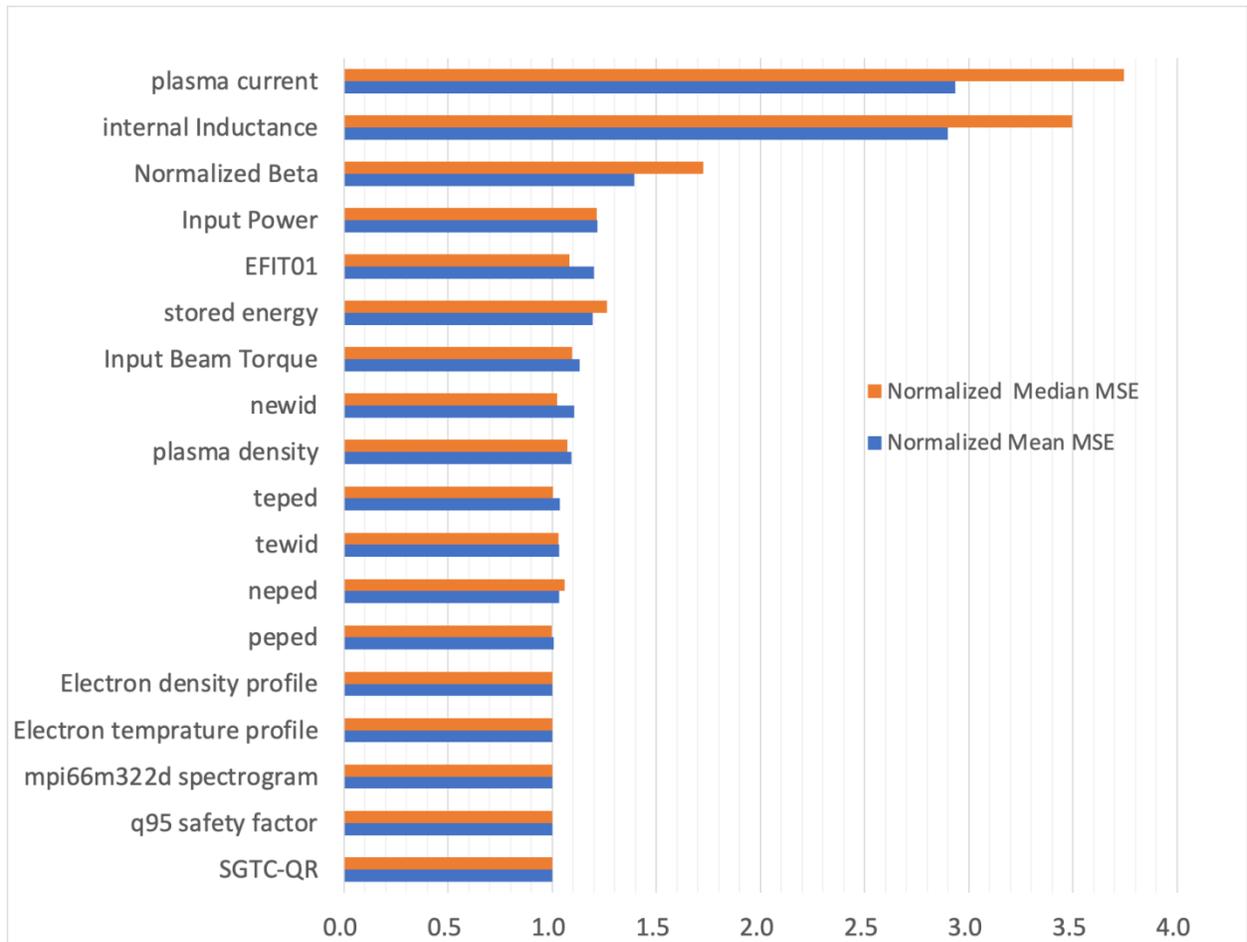

Figure 6. The sensitivity study on the testing data set. All mean-square-error data are normalized to the results of the SGTC-QR model. Each signal indicates that the signal is contaminated.

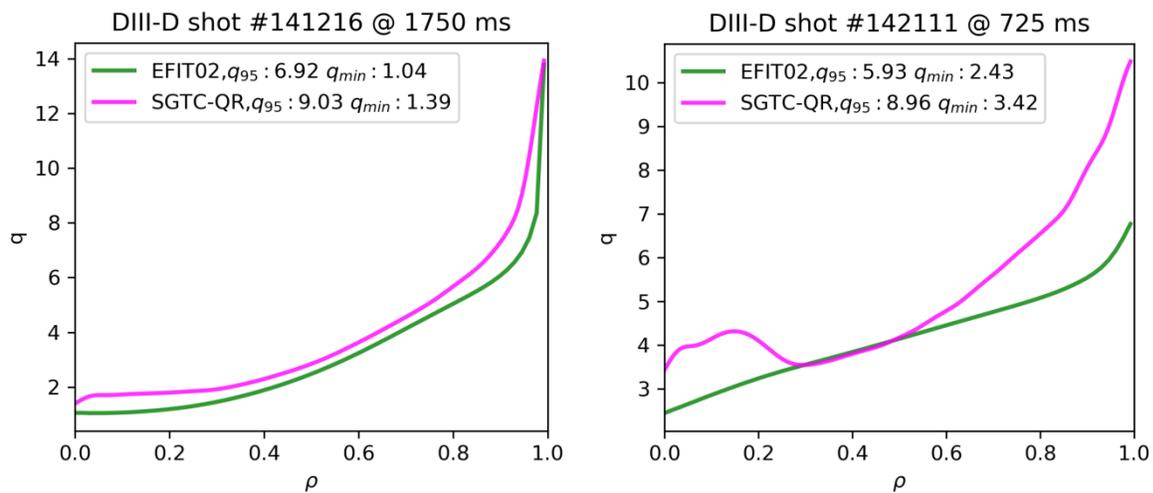

Figure 7. Comparison of q profile reconstructions for shot #141216[Brochard 2022] and # 142111 [W. Deng 2012].

## IV. Conclusions

In this work, we have designed and built a deep learning-based model for accurately reconstructing the safety factor in tokamaks. Totally 26400 equilibriums from DIII-D shots are selected for the training and testing of the model. Using the real-time 0-D and 1-D signals, the model is trained to recover the q profile from EFIT02 reconstruction. The model has demonstrated promising capability in terms of speed and accuracy. The typical prediction time is less than 1ms, and the predicted q profile is close to that from EFIT02. It shows the correlation between the input signals and the MSE constraint has been found and included during the training process. It is very helpful to use this tool to generate the q profile for the shots where MSE is unavailable. In the future, we can extend the current model to include more constraints and predict the self-consistent 2D equilibrium, pressure profile, and the current profile without solving the Grad-Shafranov equation. The SGTC-QR module will be included in the SGTC framework and the whole framework can provide abundant real-time equilibrium and perturbation information for the PCS in fusion devices.

**Reference**

[C. T. Holcomb 2006] C. T. Holcomb, et al., Rev. Sci. Instrum. **77**, 10E506 (2006)

[C. T. Holcomb 2008] C. T. Holcomb, et al., Rev. Sci. Instrum. **79**, 10F518 (2008)

[D. Spong 2012] D. Spong, et al, PHYSICS OF PLASMAS, **19**, 082511(2012)

[G. Brochard 2022] G. Brochard, et al, Nucl. Fusion **62**, 036021 (2022)

[G. Dong 2021] G. Dong, et al, Nucl. Fusion **61,** 126061 (2021)

[L. L. Lao 1985] L. L. Lao, et al, Nucl. Fusion **25**, 1611 (1985)

[L. L. Lao 1990] L. L. Lao, et al, Nucl. Fusion **30**, 1035 (1990)

[L. L. Lao 2022] L. L. Lao, et al, Plasma Phys. Control. Fusion **64**, 074001 (2022)

[J. Kates-Harbeck 2019] J. Kates-Harbeck, A. Svyatkovskiy, and W. Tang, Nature **568**, 526–531 (2019)

[S. Joung 2019] Semin Joung, et al, Nucl. Fusion **60**, 016034 (2019)

[Wróblewski 1992] D. Wróblewski and L. L. Lao, Review of Scientific Instruments **63**, 5140 (1992)

[W. Deng 2012] W. Deng, et al, Nucl. Fusion **52**, 043006 (2012)

[van Milligen 1995] B. Ph. van Milligen, et al, Neural network differential equation and plasma equilibrium solver Phys. Rev. Lett. **75,** 3594 (1995)

Acknowledgment

This work is supported by the U.S. Department of Energy (DOE) SciDAC project ISEP and used resources of the Oak Ridge Leadership Computing Facility at Oak Ridge National Laboratory (DOE Contract No. DE-AC05-00OR22725) and the National Energy Research Scientific Computing Center (DOE Contract No. DE-AC02-05CH11231). This work is partially based upon work using the DIII-D National Fusion Facility, a DOE Office of Science user facility, under Awards DE-FC02-04ER54698. This report was prepared as an account of work sponsored by an agency of the United States Government. Neither the United States Government nor any agency thereof, nor any of their employees, makes any warranty, express or implied, or assumes any legal liability or responsibility for the accuracy, completeness, or usefulness of any information, apparatus, product, or process disclosed, or represents that its use would not infringe privately owned rights. Reference herein to any specific commercial product, process, or service by trade name, trademark, manufacturer, or otherwise, does not necessarily constitute or imply its endorsement, recommendation, or favoring by the United States Government or any agency thereof. The views and opinions of authors expressed herein do not necessarily state or reflect those of the United States Government or any agency thereof.